\documentclass[twocolumn,showpacs,preprintnumbers,pra,superscriptaddress]{revtex4}
\usepackage{amsmath}
\usepackage{graphicx}
\usepackage{dcolumn}
\usepackage{bm}
\usepackage{units}
\usepackage{stmaryrd}
\usepackage[breaklinks=true,pdfborder={0 0 0}]{hyperref}
\usepackage[usenames]{color}
\usepackage{tabularx}
\usepackage{color}
\usepackage{tikz}
\usepackage{multirow}
\usepackage{amsfonts}
\usepackage{dsfont}
\usepackage{mathrsfs}
\usepackage{xcolor}
\usepackage{accents}
\usepackage{ulem}

\hypersetup{colorlinks=true,citecolor=red}

\begin{document}

\title{Seebeck Effect in Nanomagnets}
\author{Dmitry~V.~Fedorov}
\affiliation{Department of Physics and Materials Science, University of
Luxembourg, L-1511 Luxembourg City, Luxembourg}
\affiliation{Institute of Physics, Martin Luther University Halle-Wittenberg, 06099
Halle, Germany}
\affiliation{Max Planck Institute of Microstructure Physics, Weinberg 2, 06120 Halle,
Germany}
\author{Martin~Gradhand}
\email{M.Gradhand@bristol.ac.uk}
\affiliation{H.~H.~Wills Physics Laboratory, University of Bristol, Bristol BS8 1TL,
United Kingdom}
\affiliation{Institut f\"ur Physik, Johannes-Gutenberg-Universit\"at Mainz,
Staudingerweg 7, 55128 Mainz, Germany}
\author{Katarina~Tauber}
\affiliation{Institute of Physics, Martin Luther University Halle-Wittenberg, 06099
Halle, Germany}
\author{Gerrit~E.~W.~Bauer}
\affiliation{WPI-AIMR and IMR\ and CSRN, Tohoku University, Sendai, Miyagi 980-8577, Japan}
\affiliation{Zernike Institue for Advanced Materials, University of Groningen, Nijenborgh
4, 9747 AG Groningen, The Netherlands}
\author{Ingrid~Mertig}
\affiliation{Institute of Physics, Martin Luther University Halle-Wittenberg, 06099
Halle, Germany}
\affiliation{Max Planck Institute of Microstructure Physics, Weinberg 2, 06120 Halle,
Germany}
\date{\today}

\begin{abstract}
We present a theory of the Seebeck effect in nanomagnets with dimensions
smaller than the spin diffusion length, showing that the spin accumulation
generated by a temperature gradient strongly affects the thermopower. We
also identify a correction arising from the transverse temperature gradient
induced by the anomalous Ettingshausen effect and an induced spin-heat
accumulation gradient. The relevance of these effects for nanoscale magnets
is illustrated by \textit{ab initio} calculations on dilute magnetic alloys.
\end{abstract}

\maketitle

%
%\preprint{APS/123-QED}
%
%\title{Seebeck Effect in Nanoscale Ferromagnets}


%Use showkeys class option if keyword

\section{Introduction}

Spin caloritronics~\cite{Bauer2010,Bauer2012,Boona2014} addresses the
coupling between the spin and heat transport in small structures and
devices. The effects addressed so far can be categorized into several groups~%
\cite{Bauer2012}. The first group covers phenomena whose origin is not
connected to spin-orbit coupling (SOC). \textit{Nonrelativistic} spin
caloritronics in magnetic conductors addresses thermoelectric effects in
which motion of electrons in a thermal gradient drives spin transport, such
as the spin-dependent Seebeck~\cite{Slachter2010} and the reciprocal Peltier~%
\cite{Flipse2012,Goennenwein2012} effect. Another group of phenomena is
caused by SOC and belongs to \textit{relativistic} spin caloritronics~\cite%
{Bauer2012} including the anomalous~\cite{Miyasato2007} and spin~\cite%
{Cheng2008,Liu2010,Ma2010,Tauber2012,Bose2019} Nernst effects.

The Seebeck effect~\cite{Seebeck1822} or thermopower stands for the
generation of an electromotive force or gradient of the electrochemical
potential $\mu $ by temperature gradients $\nabla T$. The Seebeck
coefficient $S$ parameterized the proportionality when the charge current $j$
vanishes: 
\begin{equation}
(\nabla \mu /e)_{_{j=0}}=S\nabla T\ .  \label{Seebeck_NM}
\end{equation}%
In the two-current model for spin-polarized systems, the thermopower of a
magnetic metal reads 
\begin{equation}
S=\frac{\sigma ^{+}S^{+}+\sigma ^{-}S^{-}}{\sigma ^{+}+\sigma ^{-}}\ ,
\label{Seebeck_FM}
\end{equation}%
where $\sigma ^{\pm }$ and $S^{\pm }$ are the spin-resolved longitudinal
conductivities and Seebeck coefficients, respectively.

Here, we study the Seebeck effect in nanoscale magnets on scales equal or
less than their spin diffusion length~\cite{Comment_1} as in Figure~\ref%
{Fig.:Schematics}. Thermal baths on both sides of the sample drive a heat
current in the $x$ direction. Since no charge current flows, a thermovoltage
builds up at the sample edges that can be observed non-invasively by tunnel
junctions or scanning probes. Note that metallic contacts can detect the
thermovoltage at zero-current bias conditions, but this requires additional
modelling of the interfaces. We show in the following that in the presence
of a thermally generated spin accumulation the thermopower differs from Eq.~(%
\ref{Seebeck_FM}). We then focus on dilute ternary alloys of a Cu host with
magnetic Mn and nonmagnetic Ir impurities. By varying the alloy
concentrations we may tune to the unpolarized case $S^{+}=S^{-}$, as well as
to spin-dependent $S^{+}$ and $S^{-}$ parameters with equal or opposite
signs. The single-electron thermoelectric effects considered here can be
distinguished from collective magnon drag effects~\cite{Watzmann_2016} by
their temperature dependence.

\begin{figure}[t]
\includegraphics[width=8.5cm]{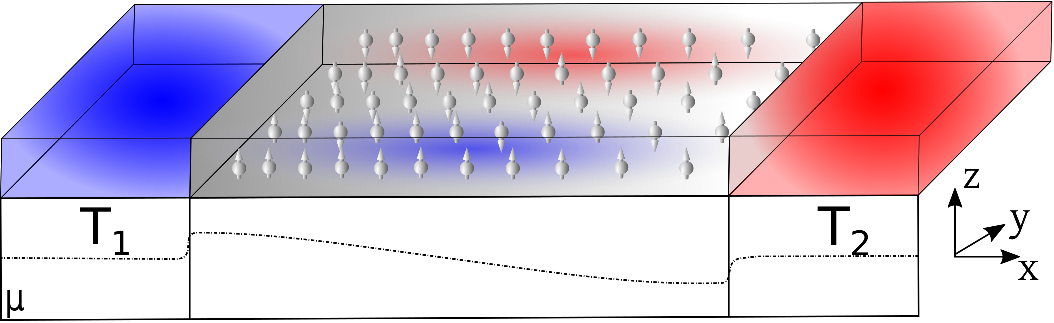}
\caption{We consider a ferromagnetic metal slab smaller than the spin
diffusion length in contact with two thermal baths hot (red) and cold (blue)
that generate a temperature gradient in the $x$ direction. The spheres with
arrows respresent the excited electrons with spin up and down parallel to
the magnetization. The thermally induced electrons are represented by their density as
well as a gradient in the grey scale of the background. The red and blue
clouds indicate transverse heat accumulation (in the $y$ direction). The dash-dotted line is the chemical potential $\protect\mu $ for a
high interface resistance to the contacts.}
\label{Fig.:Schematics}
\end{figure}

\section{Theory}

In the two-current model of spin transport in a single-domain magnet~\cite%
{Mott36,Fert68,Son87}, extended to include heat transport, the charge ($\bm{j}
$) and heat ($\bm{q}$) current densities read 
\begin{align}
\bm{j}^{\pm }& =\hat{\sigma}^{\pm }(\boldsymbol{\nabla }\mu ^{\pm }/e)-\hat{%
\sigma}^{\pm }\hat{S}^{\pm }\boldsymbol{\nabla }T^{\pm }\ ,  \label{j_pm} \\
\bm{q}^{\pm }& =\hat{\sigma}^{\pm }\hat{S}^{\pm }T(\boldsymbol{\nabla }\mu
^{\pm }/e)-\hat{\kappa}^{\pm }\boldsymbol{\nabla }T^{\pm }\ ,  \label{q_pm}
\end{align}%
where $\hat{\sigma}^{\pm }$, $\hat{S}^{\pm }$, and $\hat{\kappa}^{\pm }$ are
the spin-resolved electric conductivity, Seebeck coefficient, and heat
conductivity, respectively. All transport coefficients are tensors that
reflect crystalline symmetry and SOC. The \textquotedblleft four-current
model\textquotedblright\ Eqs.~(\ref{j_pm}) and (\ref{q_pm}) can be rewritten
as 
\begin{equation}
\begin{pmatrix}
\bm{j} \\ 
\bm{j}^{s} \\ 
\bm{q} \\ 
\bm{q}^{s}%
\end{pmatrix}%
=%
\begin{pmatrix}
\hat{\sigma} & \hat{\sigma}^{s} & \hat{\sigma}\hat{S}T & \hat{\sigma}\hat{S}%
^{s}T \\ 
\hat{\sigma}^{s} & \hat{\sigma} & \hat{\sigma}\hat{S}^{s}T & \hat{\sigma}%
\hat{S}T \\ 
\hat{\sigma}\hat{S}T & \hat{\sigma}\hat{S}^{s}T & \hat{\kappa}T & \hat{\kappa%
}^{s}T \\ 
\hat{\sigma}\hat{S}^{s}T & \hat{\sigma}\hat{S}T & \hat{\kappa}^{s}T & \hat{%
\kappa}T%
\end{pmatrix}%
\begin{pmatrix}
\boldsymbol{\nabla }\mu /e \\ 
\boldsymbol{\nabla }\mu ^{s}/2e \\ 
-\boldsymbol{\nabla }T/T \\ 
-\boldsymbol{\nabla }T^{s}/2T%
\end{pmatrix}
\label{set2}
\end{equation}%
in terms of the charge $\bm{j}=\bm{j}^{+}+\bm{j}^{-}$, spin $\bm{j}^{s}=%
\bm{j}^{+}-\bm{j}^{-}$, heat $\bm{q}=\bm{q}^{+}+\bm{q}^{-}$, and spin-heat $%
\bm{q}^{s}=\bm{q}^{+}-\bm{q}^{-}$ current densities. Here, we introduced the
conductivity tensors for charge $\hat{\sigma}=\hat{\sigma}^{+}+\hat{\sigma}%
^{-}$, spin $\hat{\sigma}^{s}=\hat{\sigma}^{+}-\hat{\sigma}^{-}$, heat $\hat{%
\kappa}=\hat{\kappa}^{+}+\hat{\kappa}^{-}$, and spin heat $\hat{\kappa}^{s}=%
\hat{\kappa}^{+}-\hat{\kappa}^{-}$. The driving forces are 
\begin{equation}
\boldsymbol{\nabla }\mu =\frac{1}{2}(\boldsymbol{\nabla }\mu ^{+}+%
\boldsymbol{\nabla }\mu ^{-})\ ,\ \boldsymbol{\nabla }T=\frac{1}{2}(%
\boldsymbol{\nabla }T^{+}+\boldsymbol{\nabla }T^{-})  \label{mu}
\end{equation}%
and the gradients of the spin $\mu ^{s}=\mu ^{+}-\mu ^{-}$~\cite%
{Son87,Johnson87,Johnson88,Valet93} and spin-heat $T^{s}=T^{+}-T^{-}$
accumulations \cite%
{Hatami2007,Heikkila2010_PRB,Heikkila2010_SSC,Bauer2012,Dejene2013,Marun2014,Wong2015}
\begin{equation}
\boldsymbol{\nabla }\mu ^{s}=\frac{1}{2}(\boldsymbol{\nabla }\mu ^{+}-%
\boldsymbol{\nabla }\mu ^{-}),\ \boldsymbol{\nabla }T^{s}=\frac{1}{2}(%
\boldsymbol{\nabla }T^{+}-\boldsymbol{\nabla }T^{-})\ .  \label{mu_s}
\end{equation}%
Finally, the tensors 
\begin{equation}
\hat{S}=\hat{\sigma}^{-1}(\hat{\sigma}^{+}\hat{S}^{+}+\hat{\sigma}^{-}\hat{S}%
^{-})  \label{S}
\end{equation}%
and 
\begin{equation}
\hat{S}^{s}=\hat{\sigma}^{-1}(\hat{\sigma}^{+}\hat{S}^{+}-\hat{\sigma}^{-}%
\hat{S}^{-})  \label{S_s}
\end{equation}%
in Eq.~(\ref{set2}) describe the charge and spin-dependent Seebeck
coefficients, respectively. In cubic systems the diagonal component $S_{ii}$, where $i$ is
the Cartesian component of the applied temperature gradient, reduces
to the scalar thermopower Eq.~(\ref{Seebeck_FM}).

\section{Results}

In the following we apply Eq.~(\ref{set2}) to the Seebeck effect in
nanoscale magnets assuming their size to be smaller than the spin diffusion
length. In this case the spin-flip scattering may be disregarded~\cite%
{Hatami2010}. We focus first on longitudinal transport and disregard $%
\boldsymbol{\nabla }T^s$. However, we also discuss transverse (Hall) effects
as well as the spin temperature gradient below. We adopt open-circuit
conditions for charge and spin transport under a temperature gradient.
Charge currents and, since we disregard spin-relaxation, spin currents
vanish everywhere in the sample: 
\begin{align}  
\label{tEq2}
0 = \hat{\sigma} (\boldsymbol{\nabla}\mu/e) + \hat{\sigma}^s (\boldsymbol{%
\nabla}\mu^s/2e) - \hat{\sigma}\hat{S} \boldsymbol{\nabla }T\ , \\
 \label{tEq3}
0 = \hat{\sigma}^s (\boldsymbol{\nabla}\mu/e) + \hat{\sigma} (\boldsymbol{%
\nabla}\mu^s/2e) - \hat{\sigma}\hat{S}^s \boldsymbol{\nabla }T\ , \\
 \label{tEq4}
\bm{q} = T \hat{\sigma} [\hat{S} (\boldsymbol{\nabla}\mu/e) + \hat{S}^s (%
\boldsymbol{\nabla}\mu^s/2e)] - \hat{\kappa} \boldsymbol{\nabla }T\ .
\end{align}

The thermopower now differs from the conventional expression given by Eq.~(%
\ref{Seebeck_FM}). Let us introduce the tensor $\hat{\Sigma}$ as 
\begin{equation}
\left. \frac{\boldsymbol{\nabla }\mu }{e}\right\vert _{_{j=0}}=\hat{\Sigma}%
\boldsymbol{\nabla }T.  \label{Seebeck}
\end{equation}%
From Eqs.~(\ref{tEq2}) and (\ref{tEq3}), we find 
\begin{equation}
\hat{\Sigma}=\left( \hat{\sigma}-\hat{\sigma}^{s}\hat{\sigma}^{-1}\hat{\sigma%
}^{s}\right) ^{-1}\left( \hat{\sigma}\hat{S}-\hat{\sigma}^{s}\hat{S}%
^{s}\right) \ .  \label{capsigma}
\end{equation}%
When the spin accumulation in Eq.~(\ref{tEq2}) vanishes we recover $\hat{%
\Sigma}\rightarrow \hat{S}$. Equation~(\ref{capsigma}) involves only directly
measurable material parameters~\cite{Comment_2}, but the physics is clearer
in the compact expression 
\begin{equation}
\hat{\Sigma}=(\hat{S}^{+}+\hat{S}^{-})/2\ .  \label{Sigma12}
\end{equation}%
The spin polarization of the Seebeck coefficient 
\begin{equation}
\left. \frac{\boldsymbol{\nabla }\mu ^{s}}{2e}\right\vert _{_{j=0}}=\hat{%
\Sigma}^{s}\boldsymbol{\nabla }T\ ,  \label{Seebeck_s}
\end{equation}%
reads 
\begin{equation}
\hat{\Sigma}^{s}=\left( \hat{\sigma}-\hat{\sigma}^{s}\hat{\sigma}^{-1}\hat{%
\sigma}^{s}\right) ^{-1}(\hat{\sigma}\hat{S}^{s}-\hat{\sigma}^{s}\hat{S})\ ,
\label{capsigma^s}
\end{equation}%
or 
\begin{equation}
\hat{\Sigma}^{s}=(\hat{S}^{+}-\hat{S}^{-})/2\ .  \label{Sigma12_s}
\end{equation}%
The diagonal elements of $\hat{\Sigma}$ govern the thermovoltage in the
direction of the temperature gradient. The off-diagonal elements of $\hat{%
\Sigma}$ represent transverse thermoelectric phenomena such as the anomalous~%
\cite{Miyasato2007} and planar~\cite{Pu2006} Nernst effects. The diagonal
and off-diagonal elements of $\hat{\Sigma}^{s}$ describe the spin-dependent
Seebeck effect~\cite{Slachter2010,Bauer2012}, as well as (also in
non-magnetic systems) the spin and planar-spin Nernst effects~\cite%
{Cheng2008,Liu2010,Ma2010,Tauber2012,Bose2019}, respectively. We do not
address here anomalous and Hall transport in the purely charge and heat
sectors of Eq.~(\ref{set2}).

\begin{table}[t]
\begin{tabular}{p{0.06\textwidth}p{0.11\textwidth}p{0.18\textwidth}p{0.1\textwidth}}
\hline\hline
System & Cu$_{0.99}$Mn$_{0.01}$ & Cu$_{0.99}$(Mn$_{0.5}$Ir$_{0.5}$)$_{0.01}$
& Cu$_{0.99}$Ir$_{0.01}$ \\ \hline
$S_{xx}^+$ & -6.87 & -7.01 & -7.09 \\ 
$S_{xx}^-$ & ~8.57 & ~1.64 & -7.09 \\ 
$S_{xx}$ & -6.14 & -4.26 & -7.09 \\ 
$\Sigma_{xx}$ & ~0.85 & -2.69 & -7.09 \\ 
$\Sigma_{xx}^s$ & -7.72 & -4.33 & ~0.00 \\ \hline\hline
\end{tabular}%
\caption{Computed spin-resolved and charge thermopowers as defined in the
text for magnetic Cu$_{0.99}$(Mn$_{0.5}$Ir$_{0.5}$)$_{0.01}$ and Cu$_{0.99}$%
Mn$_{0.01}$ as well as non-magnetic Cu$_{0.99}$Ir$_{0.01}$ dilute alloys.
The conventional spin Seebeck coefficient is shown for comparison. All
quantities are calculated at 300~K in units of $\protect\mu$V/K.}
\label{seebecktable}
\end{table}

\subsection{Longitudinal spin accumulation}

\begin{figure}[t]
\includegraphics[width=8.5cm]{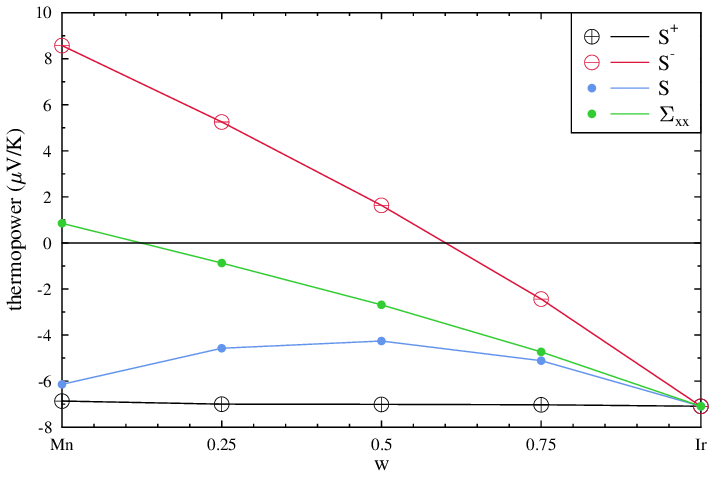}
\caption{The diagonal thermopowers $S$ and $\Sigma $, Eqs.~(\protect\ref{S})
and (\protect\ref{capsigma}), respectively, as well as the spin-resolved
thermopowers $S^{\pm }$ as calculated for dilute Cu(Mn$_{1-w}$Ir$_{w}$)
alloys at 300~K with the total impurity concentration 1~at.\%.}
\label{seebeckfig}
\end{figure}

A temperature gradient in $x$ direction $\boldsymbol{\nabla }T\parallel %
\bm{e}_{x}$ induces the voltage in the same direction: 
\begin{equation}
(\nabla _{x}\mu /e)_{_{j=0}}=\Sigma _{xx}\nabla _{x}T\ .  \label{Seebeck_xx}
\end{equation}%
In order to assess the importance of the difference between Eqs.~(\ref%
{capsigma}) and (\ref{Sigma12}) and the conventional thermopower Eq.~(\ref%
{Seebeck_FM}) we carried out first-principles transport calculations for the
ternary alloys Cu$_{1-v}$(Mn$_{1-w}$Ir$_{w}$)$_{v}$, where $w\in \lbrack
0,1]$ and the total impurity concentration is fixed to $v=1$ at.\% \cite%
{Tauber2013}. We calculate the transport properties from the solutions of
the linearized Boltzmann equation with collision terms calculated for
isolated impurities~\cite{Mertig99,Gradhand10}. We disregard spin-flip
scattering~\cite{Gradhand10}, which limits the size of the systems for which
our results hold (see below). We calculate the electronic structure of the
Cu host by the relativistic Korringa-Kohn-Rostoker method~\cite{Gradhand09}.
Figure~\ref{seebeckfig} summarizes the calculated room-temperature (charge)
thermopower Eqs.~(\ref{S}) or (\ref{capsigma}) and (\ref{Sigma12}) and their
spin-resolved counterparts, Eqs.~(\ref{capsigma^s}) and (\ref{Sigma12_s}).
Table~\ref{seebecktable} contains additional information for the binary
alloys Cu(Mn) and Cu(Ir) with $w=0$ or $w=1$ in Fig.~\ref{seebeckfig},
respectively. Here we implicitly assume an applied magnetic field that
orders all localized moments.

We observe large differences (even sign changes) between $S_{xx}^{+}$ and $%
S_{xx}^{-}$ that causes significant differences between $\Sigma
_{xx}=(S_{xx}^{+}+S_{xx}^{-})/2$ and the macroscopic $S_{xx}$. The
complicated behavior of the latter is caused by the weighting of $S^{+}$ and 
$S^{-}$ by the corresponding conductivities, see Eq.~(\ref{S}). Even though
a spin-accumulation gradient suppresses the Seebeck effect, an opposite sign
of $S_{xx}^{+}$ and $S_{xx}^{-}$ can enhance $\Sigma _{xx}^{s}$ beyond the
microscopic as well as macroscopic thermopower. Indeed, Hu \textit{et al.}~%
\cite{Hu2014} observed a spin-dependent Seebeck effect that is larger than
the charge Seebeck effect in CoFeAl. Our calculations illustrate that the
spin-dependent Seebeck effect can be engineered and maximized by doping a
host material with impurities. 

\subsection{Hall transport}

In the presence of spin-orbit interactions the applied temperature gradient $%
\boldsymbol{\nabla }T_{\mathrm{ext}}$ induces anomalous Hall currents. When
the electron-phonon coupling is weak, the spin-orbit interaction can, for
example, induces transverse temperature gradients. In a cubic magnet the
charge and spin conductivity tensors are antisymmetric. With magnetization
and spin quantization axis along $z$: 
\begin{equation}
\hat{\sigma}^{(s)}=\left( 
\begin{array}{ccc}
\sigma _{xx}^{(s)} & -\sigma _{yx}^{(s)} & 0 \\ 
\sigma _{yx}^{(s)} & \sigma _{xx}^{(s)} & 0 \\ 
0 & 0 & \sigma _{zz}^{(s)}%
\end{array}%
\right) \ ,
\end{equation}%
and analogous expressions hold for $\hat{S}$ and $\hat{S}^{s}$. A charge
current in the $x$ direction generates a transverse heat current that heats
and cools opposite edges, respectively. A transverse temperature gradient $%
\boldsymbol{\nabla }T_{\mathrm{ind}}\parallel \bm{e}_{y}$ is signature of
this anomalous Ettingshausen effect~\cite{Hu2013} gradient. From Eqs.~(\ref%
{tEq4}), (\ref{Seebeck}), and (\ref{Seebeck_s}) 
\begin{equation}
\bm{q}=\left[ T\hat{\sigma}(\hat{S}\hat{\Sigma}+\hat{S}^{s}\hat{\Sigma}^{s})-%
\hat{\kappa}\right] \boldsymbol{\nabla }T\ ,  \label{q_new}
\end{equation}%
where $\boldsymbol{\nabla }T=\bm{e}_{x}\nabla _{x}T_{\mathrm{ext}}+\bm{e}%
_{y}\nabla _{y}T_{\mathrm{ind}}$. Assuming weak electron-phonon scattering,
the heat cannot escape the electron systems and $q_{y}=0$. Equation~(\ref{q_new})
then leads to%
\begin{equation}
\nabla _{y}T_{\mathrm{ind}}=-\frac{A_{yx}}{A_{yy}}\nabla _{x}T_{\mathrm{ext}%
}\ ,  \label{T_y}
\end{equation}%
where $A_{yx}$ and $A_{yy}$ are components of the tensor 
\begin{equation}
\hat{A}=T\hat{\sigma}(\hat{S}\hat{\Sigma}+\hat{S}^{s}\hat{\Sigma}^{s})-\hat{%
\kappa}\ .  \label{Gamma}
\end{equation}%
Consequently, Eq.~(\ref{Seebeck}) leads to a correction to the thermopower 
\begin{equation}
(\nabla _{x}\mu /e)_{_{j=0}}=\left[ \Sigma _{xx}-\Sigma _{xy}\frac{A_{yx}}{%
A_{yy}}\right] \nabla _{x}T_{\mathrm{ext}}\ .  \label{nabla_x}
\end{equation}%
However, this effect should be small~\cite%
{Gradhand10_2,Lowitzer2011,Gradhand2011,Gradhand2012} for all but the
heaviest elements but may become observable when $\Sigma _{xx}$ vanishes,
which according to Fig.~\ref{seebeckfig} should occur at around $w=0.125$.

\subsection{Spin temperature gradient}

At low temperatures, the spin temperature gradient $\boldsymbol{\nabla }%
T^{s} $ may persist over length scales smaller but of the same order as the
spin accumulation~\cite{Dejene2013}. From Eqs.~(\ref{j_pm}), (\ref{Sigma12}%
), and (\ref{Sigma12_s}) it follows 
\begin{equation}
(\boldsymbol{\nabla }\mu /e)_{_{j=0}}=\hat{\Sigma}\boldsymbol{\nabla }T+\hat{%
\Sigma}^{s}\boldsymbol{\nabla }T^{s}/2\ ,  \label{LastEq1}
\end{equation}%
\begin{equation}
({\boldsymbol{\nabla }\mu }^{s}/2e)_{_{j=0}}=\hat{\Sigma}^{s}\boldsymbol{%
\nabla }T+\hat{\Sigma}\boldsymbol{\nabla }T^{s}/2\ .  \label{LastEq2}
\end{equation}%
Starting with Eq.~(\ref{set2}) and employing Eqs.~(\ref{LastEq1}) and (\ref%
{LastEq2}) for the heat and spin-heat current densities we obtain 
\begin{equation}
\bm{q}=\hat{A}\boldsymbol{\nabla }T+\hat{B}\boldsymbol{\nabla }T^{s}/2\ \ \ 
\text{and}\ \ \bm{q}^{s}=\hat{B}\boldsymbol{\nabla }T+\hat{A}\boldsymbol{%
\nabla }T^{s}/2\ ,  \label{LastEq3}
\end{equation}%
where 
\begin{equation}
\hat{B}=T\hat{\sigma}(\hat{S}\hat{\Sigma}^{s}+\hat{S}^{s}\hat{\Sigma})-\hat{%
\kappa}^{s}\ ,  \label{LastEq4}
\end{equation}%
and $\hat{A}$ is defined by Eq.~(\ref{Gamma}). With $\boldsymbol{\nabla }%
T^{s}=\bm{e}_{y}\nabla _{y}T_{\mathrm{in}}^{s}$ and $\boldsymbol{\nabla }T=%
\bm{e}_{x}\nabla _{x}T_{\mathrm{ex}}+\bm{e}_{y}\nabla _{y}T_{\mathrm{in}}$
we find 
\begin{eqnarray}
(\nabla _{x}\mu /e)_{_{j=0}} &=&\Sigma _{xx}\nabla _{x}T_{\mathrm{ex}%
}+\Sigma _{xy}\nabla _{y}T_{\mathrm{in}}+\Sigma _{xy}^{s}\nabla _{y}T_{%
\mathrm{in}}^{s}/2 \notag\\
&=&\left[ \Sigma _{xx}-\Sigma _{xy}\frac{A_{yy}A_{yx}-B_{yy}B_{yx}}{%
A_{yy}A_{yy}-B_{yy}B_{yy}}\right.  \notag \\
&&\left. -\Sigma _{xy}^{s}\frac{A_{yy}B_{yx}-B_{yy}A_{yx}}{%
A_{yy}A_{yy}-B_{yy}B_{yy}}\right] \nabla _{x}T_{\mathrm{ex}}  \label{LastEq5}
\end{eqnarray}%
assuming again $q_{y}=0$ and $q_{y}^{s}=0$. Similar to Eq.~(\ref{nabla_x}),
the Hall corrections in Eq.~(\ref{LastEq5}) should be significant only when $%
{\Sigma }_{xx}$ vanishes for $w=0.125$. However, experimentally it might be
difficult to separate the thermopowers Eq.~(\ref{LastEq5}) and Eq.~(\ref%
{nabla_x}).\newline

\subsection{Spin diffusion length and mean free path}

Our first-principles calculation are carried out for bulk dilute alloys
based on Cu and in the single site approximation of spin-conserving impurity
scattering. The Hall effects are therefore purely extrinsic. This is an
approximation that holds on length scales smaller than various spin
diffusion lengths $l_{\mathrm{sf}}$. On the other hand, the Boltzmann
equation approach is valid when the sample is larger than the elastic
scattering mean free path $l$, so our results should be directly applicable
for sample lengths $L$ that fulfill $l$ $<L\leq l_{\mathrm{sf}}.$ According
to Refs.~\onlinecite{Gradhand10_2} and \onlinecite{Gradhand2012}, for the ternary alloy Cu(Mn$%
_{0.5}$Ir$_{0.5}$) with impurity concentration of $1$~at.\% the present
results hold on length scales $26$~nm $<L\leq 60$~nm and $100$~nm $<L\leq 400
$~nm for Cu(Mn). On the other hand, for nonmagnetic Cu(Ir) the applicability
is limited to a smaller interval $10$~nm $<L\leq 16$~nm. We believe that
while the results outside these strict limits may not be quantitatively
reliable, they still give useful insights into trends.\\

\section{Summary and outlook}

In summary, we derived expressions for the thermopower valid for ordered
magnetic alloys for sample sizes that do not exceed the spin diffusion
lengths (that have to be calculated separately). We focus on dilute alloys
of Cu with Mn and Ir impurities. For 1\% ternary alloys Cu(Mn$_{1-w}$Ir$_{w}$%
) with $w<0.5$ the spin diffusion length is $l_{\mathrm{sf}}>60$~nm. In this
regime the spin and charge accumulations induced by an applied temperature
gradient strongly affect each other. By \textit{ab initio} calculations of the transport properties of  Cu(Mn$_{1-w}$Ir$_{w}$) alloys, we predict thermopowers that drastically differ from the bulk value even changing sign. Relativistic Hall effects generate spin accumulations normal to the
applied temperature gradient that become significant when the longitudinal
thermopower $\Sigma _{xx}$ vanishes, for example for Cu(Mn$_{1-w}$Ir$_{w}$)
alloys at $w\approx 0.125$. \newline

After having established the principle existence of the various corrections
to the conventional transport description it would be natural to move
forward to describe extended thin films. A first-principles version of the
Boltzmann equation including all electronic spin non-conserving scatterings
in extended films is possible, but very expensive for large $l_{\mathrm{sf}%
}. $ It would still be incomplete, since the relaxation of heat to the
lattice by electron-phonon interactions and spin-heat by electron-electron
scattering \cite{Heikkila2010_PRB,Heikkila2010_SSC} are not included. We
therefore propose to proceed pragmatically: The regime $l<l_{\mathrm{sf}}<L$
is accessible to spin-heat diffusion equations that can be parameterized by
first-principles material-dependent parameters as presented here and
relaxation lengths that may be determined otherwise, such as by fitting to
experimental results.

\section{Acknowledgements}

This work was partially supported by the Deutsche Forschungsgemeinschaft via
SFB 762 and the priority program SPP 1538 as well as JSPS Grants-in-Aid for
Scientific Research (KAKENHI Grant No. 19H00645). M.G. acknowledges
financial support from the Leverhulme Trust via an Early Career Research
Fellowship (ECF-2013-538) and a visiting professorship at the Centre for
Dynamics and Topology of the Johannes-Gutenberg-University Mainz.


\begin{thebibliography}{99}
\bibitem{Comment_2} $\hat{S}$ and $\hat{S}^s$ defined by Eqs.~(\ref{S}) and (%
\ref{S_s}) can be measured in samples much larger than the spin diffusion
length.
\end{thebibliography}
\end{document}